\documentclass{revtex4}
\usepackage{epsfig,latexsym}
\usepackage{amsmath,amscd,amssymb}

\begin{document}
\title{Theory of fluorescence correlation spectroscopy at variable observation area for two-dimensional diffusion on a meshgrid}
\author{Nicolas Destainville$^{1,2}$}

\affiliation{$^{1}$Laboratoire de Physique Th\'eorique, Universit\'e de Toulouse, CNRS, 118, route de Narbonne, 31062 Toulouse cedex, France \\
$^{2}$Institut de Pharmacologie et Biologie Structurale, Universit\'e de Toulouse, CNRS, 205, route de Narbonne, 31077 Toulouse cedex, France}

\date{\today}

\begin{abstract}
It has recently been proposed, with the help of numerical investigations, that fluorescence correlation spectroscopy at variable observation area can reveal the existence of a meshgrid of semi-permeable barriers hindering the two-dimensional 
diffusion of tagged particles, such as plasmic membrane constituents. We present a complete theory confirming and accounting for these findings.  It enables a reliable, quantitative exploitation of experimental data from which the sub-wavelength mesh size can be extracted.  Time scales at which fluorescence correlation spectroscopy must be performed experimentally are discussed in detail.
\end{abstract}

\maketitle

\section{Introduction}
The rapid development of experimental techniques for the exploration of the structure of cellular membranes at the sub-micrometric scale emphasizes the need for suitable tools to interpret the large amount of data collected. Among the many techniques routinely used in biology or biophysics laboratories, Fluorescence Correlation Microscopy (FCS) \cite{Magde72,Rigler} is very promising because it has  excellent temporal dynamics: the auto-correlation function can be built for time-scales going from nanoseconds to minutes. Even though its spatial resolution is {\em a priori} limited by optical diffraction (typically 200 nm), recent physical improvements have broken this Rayleigh limit~\cite{Webb03,Edel05}. Their spatial resolution can go down to a few tens of nanometers, thus making accessible the characterization of nano-domains  in cell plasma membranes~\cite{Wenger07}.

With this powerful tool, it is now crucial to be able to identify, in the experimental data dealing with real biological systems, the signatures of the different diffusion modes~\cite{Wenger07,Wawre05,Saxton05}. Indeed, molecules are in constant movement in live systems because the cohesion energies between biological molecules are generally of the same order of magnitude as the thermal energy $k_BT$. Thermal agitation is omnipresent at this scale and must be taken into account in the analysis of biological processes.  In particular, thermally induced diffusion certainly plays an important role in many biological functions at the molecular scale, and there is everyday new evidence that characterizing these functions cannot go without the understanding of the subtle dynamical organization of cell constituents.  As far as proteins or lipids embedded in the cellular plasma membrane are concerned, their diffusion is furthermore affected by heterogeneities of the membrane or of its close environment. For instance, they can be confined or partially confined in micro-domains. This confinement is supposed to enhance the interactions with co-confined constituents and plays a role in the biological functions of membrane molecules. The processes responsible for confinement have
been widely investigated in the last decade and several mechanisms have been proposed to account for it, depending on the biological systems~\cite{Kusumi93,Choquet03,Daumas03,Meilhac06}; see also the review~\cite{Baker07}.
Without giving greater importance to any one of these mechanisms, we focus here on the case where diffusion is hindered by linear semi-permeable obstacles (``fences"), whatever their nature, constituting a meshgrid with a typical mesh size (or mesh parameter). Cortical cytoskeleton filaments have been demonstrated to play such a role by slowing down the long-term diffusion of membrane constituents~\cite{Sheetz83,Kusumi93}, with a mesh size of several tens or few hundreds of nanometers. The diffusion appears to be confined at short times and the ensuing long-term, macroscopic diffusion coefficient, $D_M$, is much smaller than the short-term, microscopic one, $D_\mu$.

In this context and following previous works on Fluorescence Recovery After Photo-bleaching (FRAP) at variable radius~\cite{Edidin87,Edidin91,Salome98}, P.-F. Lenne and his collaborators have recently demonstrated that the use of FCS at variable observation area enables one to distinguish between several diffusion modes~\cite{Wawre05,Wenger07,Saxton05}. They proved, with the help of numerical simulations, that the three following diffusional behaviors  have a different signature: free
diffusion; diffusion on a membrane containing disjoint sub-micrometric
domains where membrane molecules are temporarily trapped and diffuse more slowly, with the idea of identifying raft-like structures; diffusion on a sub-micrometric
meshgrid constituted of joint domains, inside which the molecules
exhibit confined diffusion, but that are separated by semi-permeable barriers over
which the molecule can jump with a certain probability only, as discussed above. Our goal in the present paper is precisely to propose a rigorous analytical calculation of the two-time fluorescence autocorrelation function in the case of molecules (proteins or lipids) diffusing on such a meshgrid, thus granting the request by M.J. Saxton:  ``The theory remains to be done"~\cite{Saxton05}. We shall prove that if $t_{1/2}$ is the time at which the FCS correlation function gets reduced by a factor 2, if $w$ is the waist (defined below) of the Gaussian observation profile, then
\begin{equation}
\label{t1:2:intro}
t_{1/2} \simeq \frac{1}{4D_M} (w^2 - a^2/12),
\end{equation}
where $a$ is the mesh size. Experimentally, the measure of $t_{1/2}$ at several waists enables one to infer the value of $a$.

The two-dimensional observation profile used in FCS experiments is usually close to Gaussian~\cite{Krichevsky}, whether it is a focused laser beam or the two-dimensional section of a confocal microscope observation volume. For these reasons, we shall compute the theoretical correlation function in this Gaussian context. However, because we think that a square, uniform profile is better adapted to introduce the basic concepts useful in the calculation, we will first focus on this academic uniform case before tackling the Gaussian one. In recent works, nanometric apertures enabling the use of observation profiles much smaller than the Rayleigh limit have been developed ~\cite{Webb03,Wenger07}. \emph{A priori}, these observation profiles are closer to a uniform disc than to a Gaussian, even though the subtle optics through a nano-aperture smaller than the light wavelength deserves a more thorough analysis (see the discussion in Ref.~\cite{Webb03}). However, as it will be discussed below, the technique developed in the present paper does not allow the analysis of a uniform, circular, observation profile. We shall restrict it to square and Gaussian ones, and we shall see that our conclusions will be qualitatively equivalent in both cases. 

After revisiting in the following Section~\ref{rappels} the standard background, among which free diffusion, the central Section~\ref{mainSec} will be devoted to the main results of this paper concerning the diffusion on a meshgrid. Most numerical results obtained in previous references will be explained, including the short-time shape of the correlation functions. Finally, the last section~\ref{cl} contains concluding remarks as well as a discussion with an experimental perspective. The two appendices contain technical material useful in Section~\ref{mainSec}.

\section{Background, definitions and free diffusion}
\label{rappels}

Here we briefly recall some generalities about FCS. As justified in the Introduction, the context is the diffusion of independent particles in a quasi-two-dimensional medium (e.g. proteins or lipids in a lipidic bilayer), tagged by fluorescent labels that are observed thanks to a focused light source illuminating a small region of the sample, the area of which varies typically between a fraction of square micrometer to several ones.

In particular, we explain here how to compute the correlation function for freely diffusing particles. This simple calculation will be useful to understand more complex ones in the following sections. Furthermore, we must recover these free-diffusion laws for diffusion on a meshgrid at large time or space scales, which will be a test of the validity of our calculations. As discussed previously, we first focus on a square observation profile before considering the more physical Gaussian case. We also show how the two-dimensional problem can be written as a product of two one-dimensional ones, thus reducing our investigations to 1D diffusion.

\subsection{Correlation function}

In FCS experiments, the two-time auto-correlation function, $g^{(2)} (t)$, indicates the average degree of correlation between measures of the number of photons, $n(s)$ and $n(s+t)$ collected by the detector, at different times $s$ and $s+t$~\cite{Magde72,Rigler}. The decorrelation depends on the diffusion coefficient(s) of the tagged particles. Each measure of $n$ is a number of photons detected per sampling time $\Delta t$. More precisely, the correlation function is defined as follows:
\begin{equation}
g^{(2)} (t)= \frac{\langle n(s) n(s+t) \rangle}{\langle n(s)\rangle^2}.
\label{g2}
\end{equation}
The brackets mean an average over a large amount of realizations of the system (\emph{ensemble} average). Because of translational symmetry in $s$ at equilibrium, the previous quantity is independent of $s$. However, owing to the ergodic theorem, it can also be seen as an average over successive times $s$ in a same, very long trajectory (\emph{temporal} average). When needed, we shall adopt the latter point of view in the following.

To detect fluorophores in the observed area, an excitation light is used, with a profile $I(\mathbf{r})=I(x,y)$ depending on the two-dimensional position $\mathbf{r}=(x,y)$ on the sample. Then the number of collected photons at time $s$ is
\begin{equation}
\label{n:s}
n(s) = \Delta t \int \mathrm{d}x\mathrm{d}y \; I(x,y) Q \rho(x,y,s),
\end{equation}
where the numerical factor Q is the product of the absorption cross section of one fluorophore by the fluorescence quantum yield and the fluorescence efficiency~\cite{Krichevsky,Rigler} and $\rho(x,y,s)$ is the number of fluorophores per unit area at time $s$, whether they are observed or not during $\Delta t$: $\rho$ depends on sample and realization only, and not on the illumination profile. A straightforward calculation~\cite{Rigler} shows that at the thermodynamic limit, $g^{(2)} (t)$ has the typical form:
\begin{equation}
\label{g2bis}
g^{(2)} (t) = 1 + \frac{1}{S_{\mathrm{eff}}\langle \rho \rangle} C_{I(\mathbf{r})}(t).
\end{equation}
In this expression, $\langle \rho \rangle$ is the ensemble average of the density of fluorophores in the system. From now, we assume without loss of generality that 
$\langle \rho \rangle =1$. Furthermore, $S_{\mathrm{eff}}$ is an effective observation area:
\begin{equation}
\label{Seff} 
S_{\mathrm{eff}} \equiv \frac{\left[ \int \mathrm{d}x \mathrm{d}y \; I(x,y) \right]^2}{\int \mathrm{d}x \mathrm{d}y \; \left[ I(x,y) \right]^2}.
\end{equation}
As for $C_{I(\mathbf{r})}(t)$, it is the core of the auto-correlation function. With our choice of normalization by $S_{\mathrm{eff}} \langle \rho \rangle$, it decreases from 1 at $t=0$ to 0 at large times. Its precise form depends on the observation profile $I(\mathbf{r})$. The prefactor $1/(S_{\mathrm{eff}}\langle \rho \rangle)$ is equal to the usual one, $1/\bar N$, found in the literature~\cite{Wawre05,Wenger07,Rigler,Krichevsky}.

At the experimental level, the measure of $g^{(2)} (t)$ on time-scales ranging from nanoseconds to minutes provides a very good estimate of $C_{I(\mathbf{r})} (t)$.
An important quantity that is extracted from FCS data is the diffusion or decorrelation time,  $t_{1/2}$, defined by
\begin{equation}
\label{diff:time}
C_{I(\mathbf{r})} (t_{1/2}) = \frac{1}{2}.
\end{equation}
By a scaling argument, one can already foresee that $t_{1/2}$ will be proportional to the ratio of the observation area and a diffusion constant. The aim of the present work is to precise this rough argument and to explicit what additional information can be extracted from FCS experiments.

\subsection{Normalized observation profiles}
\label{profiles}

As justified in the introduction, we study in this paper two different two-dimensional illumination intensity profiles (observation profiles) $I(x,y)$: a square, uniform one and a Gaussian one. We normalize them by $\int I(x,y) \; \mathrm{d}x \mathrm{d}y   =1$, without loss of generality since the normalization disappears in Eq.~(\ref{g2}). The square profile is
\begin{equation}
\label{profil:carre}
I(x,y)  =  \frac{1}{L^2}  \mathbf{1}_{[-L/2,L/2]} (x) \mathbf{1}_{[-L/2,L/2]} (y),
\end{equation}
where $\mathbf{1}_{[-L/2,L/2]}$ is the characteristic function of the segment $[-L/2,L/2]$, equal to 1 on the segment and to 0 anywhere else. One calculates that $S_{\mathrm{eff}} = L^2$. As for the normalized Gaussian profile,
\begin{equation}
\label{profil:gauss }
I(x,y) =  \frac{2}{\pi w^2}\exp\left(  -\frac{2(x^2+y^2)}{w^2} \right),
\end{equation}
where $w$ is the profile waist (twice its mean standard deviation $\langle x^2 \rangle^{1/2}$). Furthermore $S_{\mathrm{eff}} =  \pi w^2$.

\subsection{From two to one dimension}

In this paper, we focus on diffusion in two dimensions. But it appears that the problem can be reduced to a one-dimensional one because: (i) the square meshgrid is a product of two one-dimensional grids of mesh parameter $a$; (ii) the diffusion operator~-- the Laplacian in the free-diffusion case, or a more complex Fokker-Planck one~\cite{Risken} in a meshgrid, see below~-- is a tensor product of two one-dimensional diffusion operators; and (iii) the observation profile can be written as a product of two one-dimensional ones, both in the square and in the Gaussian cases: 
\begin{equation}
I(x,y) = I_1(x) I_2(y),
\end{equation}
where $I_1(x)$ and $I_2(y)$ are normalized. If the observation profile is uniform on a square, then $I_1(x)$ and $I_2(y)$ are proportional to the characteristic function of a segment of length $L$. Such $I_1(x)$ or $I_2(y)$ will be referred to as ``uniform profiles" in the following. If the two-dimensional profile is Gaussian, then $I_1(x)$ and $I_2(y)$ are also Gaussian, with the same waist as $I(x,y)$. To simplify, $I_1(x)$ will also be written $I(x)$ when this notation is not ambiguous.

In these conditions, it is proven~\cite{Krichevsky} that the correlation function $C_{I(\mathbf{r})}$ above is nothing but the product of two one-dimensional correlation functions $C(t)$, also depending on $L$ or $w$. The initial problem reduces to the study of one-dimensional diffusion on a lattice of equally spaced punctual barriers that can be passed with a certain probability. In the cases under interest here, the observation profile is either uniform of length $L$ or Gaussian of waist $w$. The only difference is that the decorrelation time $t_{1/2}$, a function of $L$ or $w$, is now the solution of 
\begin{equation}
\label{deft1:2}
C(t_{1/2})=\frac{1}{\sqrt{2}}. 
\end{equation}
We must also take care of tracking correctly the prefactor $1/S_{\mathrm{eff}}$ coming from our choice of normalization of $C_{I(\mathbf{r})}$ (the case of the factor $1/\langle \rho \rangle$ has been fixed by setting it to 1): a prefactor $1/\sqrt{S_{\mathrm{eff}}}$ is distributed to each one-dimensional $C$ (see below).

Note that we could factorize in the same way a two-dimensional problem where the grid has a two different mesh parameters $a_x$ and $a_y$; or where the observation profile has two different characteristic lengths in the $x$ and $y$ directions ($L_x$ and $L_y$ or $w_x$ and $w_y$). For simplicity's sake, we shall focus on the ``isotropic'' case.

By contrast, a circular observation profile cannot be written as a product of one-dimensional profiles. It is for this reason that this case cannot be addressed by using the tools proposed in the present paper. Note also that the previous formal decomposition in a product of one-dimensional problems can be interesting from a numerical perspective~\cite{Wawre05} since it reduces the computational complexity.

\subsection{Calculation of the free-diffusion correlation function}
\label{free}

Here we give a rapid summary of calculations that can be found in former references dealing with FCS, \emph{e.g.} Refs.~\cite{Magde72,Rigler}. In this special case, the function $C$, the diffusion coefficient and the decorrelation time $t_{1/2}$ are denoted respectively by $\Gamma$, $\Omega$ and $t_0$ (thus $\Gamma(t_0)=1/\sqrt{2}$). We also introduce the adimensional  variable $x$:
\begin{equation}
x = \frac{2\sqrt{\Omega t}}{L} \qquad (\mathrm{resp. }  \ x = \frac{2\sqrt{\Omega t}}{w})
\end{equation}
for a uniform (resp. Gaussian) observation profile. Since, by scaling arguments, $t_0$ will be of order $L^2/\Omega$ or $w^2/\Omega$, the value of $x$ corresponding to $t_0$ (denoted by $x_0$) will be of order 1 in time intervals of interest.

\begin{figure}[ht]
\begin{center}
\ \psfig{figure=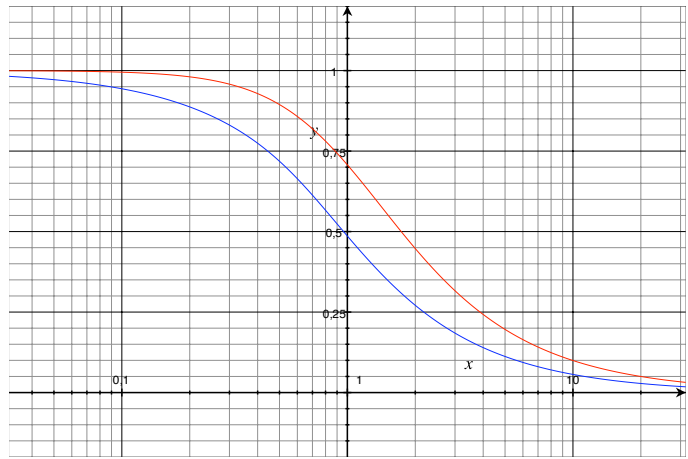,height=6cm} \
\end{center}
\caption{Fluorescence correlations in dimension 1:  Graphs of the functions $\Gamma$, as defined in Eqs.~(\ref{C:x:def:square}) and (\ref{C:x:def:gauss}), for the uniform (upper curve) and Gaussian (lower curve) observation profiles, in log-linear coordinates. The reduced variable $x$ is defined as $x=2\sqrt{D_M t}/{L}$ in the uniform case or $x=2\sqrt{D_M t}/{w}$ in the Gaussian one. Both functions essentially differ by an abscissa-shift.}
\label{C:x:courbe}
\end{figure}

If $\hat I(q)=1/\sqrt{2\pi}\int \exp(-iqx) I(x) \mathrm{d}q $ is the Fourier transform of $I(x)$, then the correlation function $\Gamma(t)$ depends on the observation profile as follows~\cite{Krichevsky}:
\begin{equation}
\label{C:I }
\Gamma(t) = \sqrt{S_{\mathrm{eff}}}\int_{-\infty}^{\infty} \mathrm{d}q \; | \hat I(q) |^2 e^{-\Omega q^2 t} .
\end{equation} 
In this equation, the only contribution of the underlying free diffusion is contained in $\Omega$. Note that $\Gamma(0)=1$, as required, by definition of $S_{\mathrm{eff}}$ and because $I$ is normalized. For a uniform observation profile, one gets after integration
\begin{equation}
\Gamma(x) = {\rm erf}(1/x)-\frac{x}{\sqrt{\pi}} (1 -
\exp(-1/x^2)),
\label{C:x:def:square}
\end{equation}
where ${\rm erf}$ is the usual error function~\cite{Abra}. The graph of $\Gamma$ is displayed in Figure~\ref{C:x:courbe}. By definition, $x_0$ is defined by $\Gamma(x_0) = 1/\sqrt{2}$. The unique solution is $x_0 \simeq 0.520$, a transcendental real. At the leading order in $L$,
$t_{1/2} = \frac{x_0^2}{4 \Omega} L^2$. This means that the $t_{1/2}$ vs $L^2$ 
profiles are strictly linear with slope $x_0^2/(4 \Omega)$.

For a Gaussian observation profile, one gets the following function, the graph of which is also given in Figure~\ref{C:x:courbe}: $\Gamma(x) = 1/\sqrt{1+x^2}$. In this case $x_0=1$, a much simpler solution, and the decorrelation time obeys $t_{1/2} = L^2/(4 \Omega)$, and $\Gamma(t)$ can be written
\begin{equation}
\Gamma(t) = \frac{1}{\sqrt{1+t^2/t_{1/2}^2}}.
\label{C:x:def:gauss}
\end{equation}

Since $x_0 \simeq 0.520 \approx 1/2$ in the uniform case, $L/2$ and $w$ play essentially the same role above, which means that the waist $w$ can be compared to the half-width of a uniform observation profile for FCS purposes.

\section{Diffusion on a meshgrid}
\label{mainSec}

\begin{figure}[ht]
\begin{center}
\ \psfig{figure=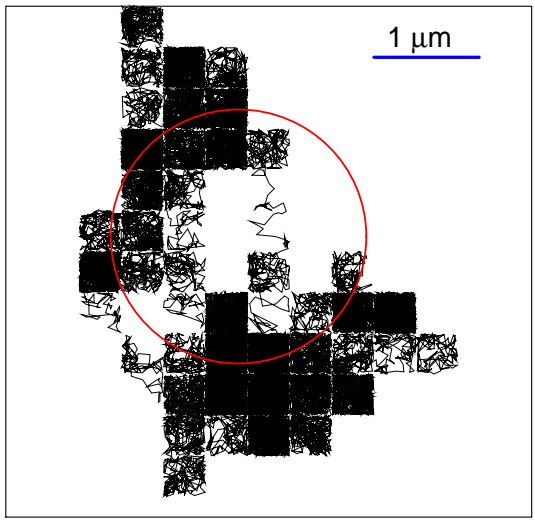,width=7cm} \
\end{center}
\caption{The simulated trajectory of one particle diffusing in a regular two-dimensional meshgrid of mesh parameter $a=400$~nm. It diffuses freely in each square domain, with diffusion coefficient arbitrarily set to $D_\mu=0.06$~$\mu$m$^2$s$^{-1}$, and it is generally subject to reflecting boundary conditions on the square edges. However, it can jump over edges with a low probability, thus resulting in an apparent slow diffusion at large times, with diffusion coefficient $D_M=D_\mu/15$ here. The circle materializes the observation area, in which the number of observed particles fluctuates. The correlation function $g^{(2)}$ measures these fluctuations and their typical time scales.}
\label{mesh:ex}
\end{figure}

Here we consider a two-dimensional system still constituted of independent particles,  but the diffusion of which is now hindered by a meshgrid of linear obstacles. For simplicity~\cite{Wawre05}, we suppose that the meshgrid is square, with mesh parameter $a$, as illustrated in Fig.~\ref{mesh:ex}. The edges of the meshgrid are semi-permeable walls: they generally prevent the motion of diffusers, resulting in free diffusion with reflecting boundary conditions in square regions. We denote by $D_\mu$ the associated (microscopic) diffusion coefficient. But the particles can also jump over these obstacles with a low probability (e.g. by thermal activation).  In the following, the squares constituting the meshgrid will be referred to as ``domains" or ``boxes", in which the diffusion is confined at intermediate times. At larger times, we expect a slow diffusion mode, with a (macroscopic) diffusion coefficient, denoted by $D_M$, and calculated below. It should be governed by the laws derived in the previous section, where the free diffusion coefficient $\Omega$ has to be replaced by $D_M$. We suppose that time-scales are well separated, as it is usually required in such a context~\cite{Wenger07,Wawre05,Daumas03,Meilhac06,Salome98}: $D_M \ll D_\mu$. This means that the barriers are true (semi-permeable) obstacles, that have a chance to be identified as such experimentally.

A question that arises at this stage is that of the position of the observation profile relatively to the meshgrid. For example, it has already been observed~\cite{Wawre05,Wenger07} that FCS profiles differ according to whether the center of the Gaussian profile falls close to a vertex or near the center of a domain. This issue will be carefully questioned in the following. Experimentally, it is common to perform many experiments on the same (model or live) system before averaging observables. This procedure is used to measure $t_{1/2}$. However, the averaging process can be performed at two different steps,
leading \emph{a priori} to slightly different values of $t_{1/2}$. Either different correlation functions $C(t,w)$ are measured before averaging them. The resulting average is denoted by $\bar C(t,w)$ and $t_{1/2}$ is defined by $\bar C(t_{1/2})=1/2$. Or one $t_{1/2}$ is measured for each experiment and the final value of $t_{1/2}$ is their average over realizations. In the following calculations, we shall adapt the first point of view, for the sake of simplicity. However, we shall prove in the last section devoted to discussions that both procedure are equivalent for experimental determination of the mesh parameter $a$, the main purpose of the present work.

\subsection{Model, box approximation and time scales}

\begin{figure}[ht]
\begin{center}
\ \psfig{figure=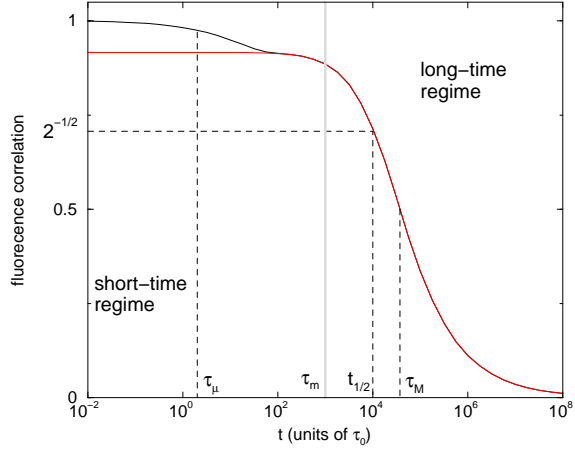,height=6cm} \
\end{center}
\caption{Example of fluorescence correlation function $\bar C(t,L)$ for a uniform observation profile of length $L=4$, after integration on the position of the profile center relatively to the meshgrid of mesh parameter $a$. The time unit $\tau_0$ is arbitrary. The diffusion parameters are $D_M = 10^{-4}a^2 \tau_0^{-1}$ and $D_\mu = 5.10^{-2}a^2 \tau_0^{-1}$. The black curve shows the entire profile whereas the red one displays the long-term component only. The ``microscopic" time-scale $\tau_\mu \sim 1 \;  \tau_0$ and the ``macroscopic" one $\tau_M \sim 10^5\;  \tau_0$ are clearly identifiable, corresponding to the two successive fall-offs in $\bar C$. The ``mesoscopic" time-scale $\tau_m$  in-between (gray vertical bar) separates the short-time regime on the left and the long-time one on the right. The decorrelation time $t_{1/2}$ is the unique solution of $\bar C(t_{1/2},L)=1/\sqrt{2}$. The knowledge of the long-term component suffices if one wants to calculate $t_{1/2}$.}
\label{ex:fcs}
\end{figure}

As it was justified above, we now focus on the one-dimensional case, that is to say the diffusion on a line with regularly spaced semi-permeable punctual obstacles. Anticipating on the following, we refer to Figure~\ref{ex:fcs} where an example of correlation function $\bar C(t,L)$ is displayed (upper curve). It corresponds to a uniform observation profile but it also qualitatively matches the Gaussian case. One clearly identifies two
time-scales in this graph, corresponding to the two successive fall-offs in the correlation function. We shall see that the first (small) fall-off, around $\tau_\mu$, is associated with the ``microscopic" decorrelation \emph{inside} domains on the border of the observation area; The second (large) one, around $\tau_M$, accounts for the ``macroscopic" decorrelation inside the whole observation area.

In order to compute the correlation function $C(t)$, we shall also need to
separate the slow and rapid time scales in the averaging process.
Having already identified the two typical ``microscopic" and ``macroscopic"
time scales $\tau_\mu$ and $\tau_M$, we introduce an additional intermediate ``mesoscopic" time scale, $\tau_m$, such that $\tau_\mu \ll \tau_m \ll \tau_M$.  When averaging over times $t$ in a trajectory below, we shall distinguish between
short-term averages, denoted by $\langle \cdot \rangle_{st}$ and
long-term ones, denoted by $\langle \cdot \rangle_{lt}$. This means
that we first calculate averages on segments $[0,\tau_m]$,
$[\tau_m,2\tau_m]$, etc\ldots, then we take the mean value of these
short-term averages. Formally, in Eq.~(\ref{g2}), $\langle \cdot
\rangle=\langle \langle \cdot \rangle_{st} \rangle_{lt}$. In
other words, in the reciprocal space, $\langle \cdot \rangle_{st}$
(resp. $\langle \cdot \rangle_{lt}$) means integration over
frequencies larger (resp. smaller) than $1/\tau_m$.

\bigskip

\bigskip

\begin{figure}[ht]
\begin{center}
\ \psfig{figure=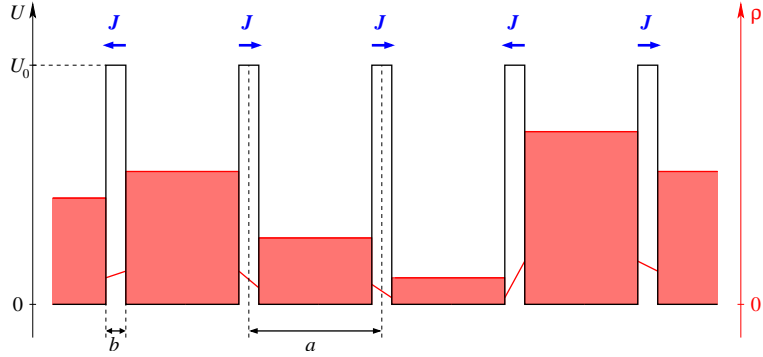,width=10cm} \
\end{center}
\caption{Quasi-static box approximation in one dimension. The boxes of width $a$ are separated by potential barriers (in black) of width $b \ll a$ and height $U_0 \gg k_BT$. The particle density $\rho$ (in red) is quasi-uniform in domains because the exchange rate between domains is much slower than the equilibration time inside domains. The difference of filling between adjacent domains results in a weak density current, $\mathbf{J}=-D_{\mu }\mathbf{\nabla} \rho$, at the barrier, due to the gradient of $\rho$ inside the barrier. This is the cause of the slow long-term diffusion.}
\label{boxes}
\end{figure}

The model used in this paper to tackle the diffusion on a meshgrid is schematized in 
Figure~\ref{boxes}. Particles diffuse in a periodic potential $U(x)$, of period the mesh parameter $a$. It is constituted of wells separated by barriers of width $b\ll a$ and of finite height $U_0$ (``punctual" semi-permeable obstacles). The fact that barriers are semi-permeable implies that $U_0 \gg k_BT$ so that jumps over barriers, requiring an energy $U_0$, are rare. This argument will be refined below. The particle density $\rho(x,t)$ is a variable indicating the probability, $\rho(x,t) \; \mathrm{d}x$, of finding a particle at time $t$ in the interval $[x,x+\mathrm{d}x]$. By definition, at equilibrium, $\langle \rho(x,t) \rangle = \langle \rho \rangle = 1$, in conformity with our
previous prescription. 

For the sake of generality, we can set two different diffusion coefficients in wells and barriers, denoted respectively by $D_\mu$ and $D_b$. The equation governing the temporal evolution of $\rho$ is then the Fokker-Planck equation~\cite{Risken}. 
This equation can be solved in the present case. However, because this generalization does not bring any insight into the physics and because it needlessly complicates the mathematical formalism, we suppose from now that $D_b=D_\mu$. The Fokker-Planck equation then amounts to the simpler Smoluchowski one~\cite{Risken}. With our choice of $U(x)$, the situation simplifies again because a classical diffusion equation is valid inside wells or barriers, where both the potential, $U$, and the diffusion coefficient, $D_\mu$, are constant. But it is not able to tackle the variations of $U$ with $x$. In the special case under interest here, the Smoluchowski equation simplifies into the ``jump condition"~\cite{Risken} because there exist only isolated discontinuities of $U$: if the discontinuity occurs at $x=x_0$, and if $\Delta U=U(x_0+\varepsilon)-U(x_0-\varepsilon)$, then
\begin{equation}
\label{jump:cond}
\rho(x_0+\varepsilon)= \exp \left( - \frac{\Delta U}{k_B T}  \right)  \rho(x_0-\varepsilon).
\end{equation}
Since $U_0 \ll k_BT$, $\rho$ is much smaller inside barriers than outside, as illustrated in Figure~\ref{boxes}.

The principle of the \emph{quasi-static box approximation} scheme developed here is also exemplified in this figure. The equilibration (or decorrelation) time-scale inside a well, denoted above by $\tau_\mu$, is given by the resolution of the diffusion equation in a square well, the boundaries of which can be considered as perfectly reflecting since jumps are scarce. One gets (see Section~\ref{short:uniform}):
\begin{equation}
\label{tau:mu}
\tau_\mu = \frac{a^2}{\pi^2 D_\mu}.
\end{equation}
This means that a non-uniform distribution $\rho$ inside the well will equilibrate and become uniform after a time much larger than $\tau_\mu$. For the example given in Figure~\ref{ex:fcs}, on gets $\tau_\mu \simeq 2.0$, as anticipated. 

Now we focus on what happens inside the barriers. Since $b \ll a$ and particle densities are much smaller inside barriers (Eq.~(\ref{jump:cond})), the probability of finding a particle inside a barrier is very small as compared to the wells. Wells behave like reservoirs with regard to barriers and impose boundary conditions to $\rho$ in barriers. In addition, the equilibration time inside barriers, $\tau_b$ is also given by Eq.~(\ref{tau:mu}): $\tau_b = b^2/(\pi^2 D_\mu)$. Then the diffusion equation becomes $\partial^2 \rho / \partial x^2 = 1/D_\mu \; \partial \rho / \partial t =  0$ (equilibrium condition), that is to say $\rho(x)= A x + B$. Thus $\rho$ is affine in barriers, its values on the boundaries being fixed by the filling of both adjacent boxes (see Figure~\ref{boxes}). The quasi-static approximation assumes that all time-scales in the problem are much larger than $\tau_\mu$ and $\tau_e$ (which is checked \emph{a posteriori} for self-consistency), so that the $\rho$ profiles inside boxes and barriers can be considered at equilibrium. They can vary with time, but very slowly. 

Indeed, suppose that two adjacent boxes of indices $i$ and $i+1$ are differently filled, as in Figure~\ref{boxes}. Then a particle current appears through the barrier from the more filled box towards the less filled one. This current is denoted by $\mathbf{J}_{i \rightarrow i+1}$. It is positive if particles go from $i$ to $i+1$, and negative in the converse case. More quantitatively, the total particle number inside the box of index $i$ at time $t$, $\int_{\mathrm{box}} \rho(x,t) \; \mathrm{d}x$, is denoted by $W_i(t)$. The particle density on the box boundary is therefore $W_i(t)/a$. The condition on the left-hand (resp. right-hand) boundary of the barrier between boxes $i$ and $i+1$ is thus $\rho=W_i(t)/a \;  \exp(-U_0/k_BT)$ (resp. $\rho=W_{i+1}(t)/a \;  \exp(-U_0/k_BT)$). Since $\rho$ is affine inside the barrier, $\nabla \rho = \exp(-U_0/k_BT) (W_{i+1}(t)-W_{i}(t))/(ab)$ and 
\begin{equation}
\label{J}
\mathbf{J}_{i \rightarrow i+1} = - D_\mu \nabla \rho = -D_\mu \exp(-U_0/k_BT) \; \frac{W_{i+1}(t)-W_{i}(t)}{ab}.
\end{equation}
In the following section, we shall write and solve the diffusion equation satisfied by the $W_i(t)$. It will ensue from the previous expression of $\mathbf{J}$.

To conclude this section, we have demonstrated that, in the conditions under interest in this paper, the diffusion on a meshgrid can be treated in the quasi-static box approximation, where domain fillings, $W_i(t)$, are uniform and vary slowly. With respect to the short-term and long-term averages discussed above, it means that the short-term average $\langle \cdot \rangle_{st}$ is implicitly already realized when supposing that $\rho(x,t) = W_i(t)$ in box $i$, since $\rho$ does not display short-term fluctuations any-longer. Again implicitly, all remaining averages $\langle \cdot \rangle$ are in fact long-term ones, $\langle \cdot \rangle_{lt}$. Finally, we remark that because $b \ll a$ and particle densities are small inside barriers, the probability of finding a particle inside a barrier is very small and will be neglected.

\subsection{Discrete diffusion equation and box-box two-time correlation function}

The evolution of $W_i(t)$ ensues from the currents $\mathbf{J}_{i \rightarrow i+1}$ and $\mathbf{J}_{i \rightarrow i-1}$:
\begin{equation}
\label{discrete:diff}
\frac{\mathrm{d}}{\mathrm{d}t} W_i(t) = -(\mathbf{J}_{i \rightarrow i+1} + \mathbf{J}_{i \rightarrow i-1}) = D_M \; \frac{W_{i-1}(t) - 2W_i(t) + W_{i+1}(t)}{a^2}.
\end{equation}
This is a discrete-position diffusion equation, the right-hand-side fraction being nothing but a discrete Laplacian on a one-dimension lattice of lattice-spacing $a$. Therefore, as anticipated, the long-term behavior of particles is diffusional, with the long-term diffusion coefficient 
\begin{equation}
\label{Omega}
D_M = \frac{a}{b} D_\mu e^ { - \frac{U_0}{k_B T}}. 
\end{equation}
The three parameters $b$, $D_\mu$, and $U_0$ eventually amount here to a single one, $D_M$. Note that the condition $D_M \ll D_\mu$ required above imposes an additional condition on $b$: $a \gg b \gg a \exp(-U_0/k_BT)$. In practice, $U_0 > 5 k_B T$ suffices to ensure that such a  value of $b$ can exist: $b \sim a/10$ satisfies both conditions. 

A new box-exchange time-scale appears in this equation, 
\begin{equation}
\label{tau:e}
\tau_e = \frac{a^2}{D_M} = \frac{ab}{D_\mu}\; e^ { \frac{U_0}{k_B T}},
\end{equation}
that indicates at which rate two differently filled adjacent boxes equilibrate. The condition $\tau_e \gg \tau_\mu$, that reads $b \gg a/\pi^2 \exp(-U_0/k_BT)$, is equivalent to the previous one $D_M \ll D_\mu$. The second condition, $\tau_e \gg \tau_b$, being equivalent to $a/b \gg \exp(-U_0/k_BT)/\pi^2$, is automatically satisfied.

The discrete diffusion equation~(\ref{discrete:diff}) is solved by introducing its Green's function: if $W_i(t=0) = \delta_{i,0}$ (Kronecker symbol), then $W_i(t)  =I_i (2 \tilde D_M t) \exp(-2 \tilde D_M t)$, with $\tilde D_M = D_M/a^2$, can be checked to be the (unique) solution~\cite{Abra}. In particular, owing to the properties of the modified Bessel functions of the first kind $I_i(z)$, $ W_i(t)  \simeq (4 \pi \tilde D_M t)^{-1/2}$ at large $t$, for any fixed $i$. This large $t$ behavior will be refined in the next section.

In the following, we shall also need the box-box, two-time correlation functions of the $W_i(t)$:
\begin{equation}
\label{correl:fct}
\gamma_{ij} (t) = \langle W_i(s) W_j(s+t) \rangle 
- \langle W_i(s)\rangle  \langle W_j(s+t) \rangle.
\end{equation}
Here and in the following section, the averages $\langle \cdot \rangle$ must be understood as the long-term ones, $\langle \cdot \rangle_{lt}$.
This correlator can be computed thanks to the following argument: $\mathrm{d}/\mathrm{d}t \; \gamma_{ij} (t) = \langle W_i(s) \; \mathrm{d}/\mathrm{d}t \; W_j(s+t) \rangle$, and Eq.~(\ref{discrete:diff}) can be injected in this relation, which shows that the $\gamma_{ij}$ obey the same diffusion equation as the $W_i$. We recall that $\langle W \rangle =1$. Thus $\gamma_{ij} (t=0) = \delta_{ij}$, and it ensues that
\begin{equation}
\label{correl:fct2}
\gamma_{ij} (t) =  I_{j-i} (2 \tilde D_M t) \exp(-2 \tilde D_M t).
\end{equation}
This relation will be useful to calculate $C(t)$ below.

\subsection{Calculation of the long-term correlation function for a uniform observation profile}
\label{square:long}

\begin{figure}[ht]
\begin{center}
\ \epsfig{figure=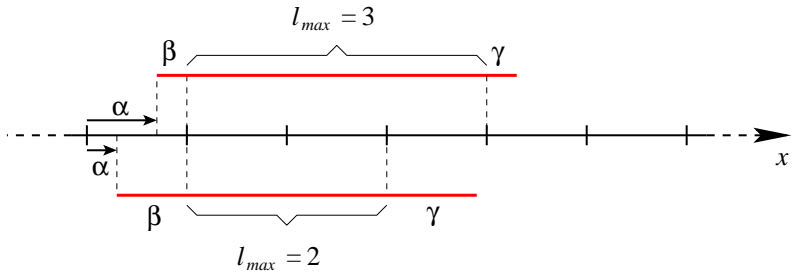,width=10.5cm} \
\end{center}
\caption{Definition of the various lengths used in the calculation of Section~\ref{square:long}. The short vertical lines indicate the positions of the semi-permeable barriers. All lengths are in units of the mesh parameter $a$. In this example, the observation segments (in red) have length $L=3.6$. The number of boxes that it covers entirely is denoted by $l_{max}$. Depending on its position $\alpha \in [0,1[$ relatively to the meshgrid, $l_{max}$ can be equal to $\ell=3$ (top) or $\ell-1=2$ 
(bottom). The residues $\beta$ and $\gamma$ give the overlap of the observation segment with the first and last observed boxes.}
\label{defs}
\end{figure}

We first consider the correlation function at large times: $t > \tau_m$. As it will be confirmed in the ensuing subsection (devoted to short times, $t < \tau_m$), boxes can be considered at equilibrium in the long-term regime because local box fluctuations are then decorrelated. The normalized observation profile is $I(x) = \frac{1}{L} \mathbf{1}_{[-L/2,L/2]}$. The one-dimensional correlation function $C(t,L,\alpha)$ depends obviously on $t$, but also on the profile width $L$ (it is the core of the present work) and on the position $\alpha$ of this profile relatively to the meshgrid (see Figure~\ref{defs}). By definition, and after some algebraic manipulation using in particular the correlators $\gamma_{ij}$,
\begin{eqnarray}
C(t,L,\alpha) & = & \frac{1}{L} \langle
\sum_{i=-\infty}^{\infty} [W_i(s) - \langle W_i \rangle]
\int_{ia}^{(i+1)a} \mathbf{1}_{[-L/2,L/2]}(x-\alpha) \; {\rm d}x
 \nonumber \\
 & & \times \sum_{j=-\infty}^{\infty} [W_j(s+t) - \langle W_j \rangle]
\int_{ja}^{(j+1)a} \mathbf{1}_{[-L/2,L/2]}(y-\alpha) \; {\rm d}y
\rangle
\\
 & = & \frac{1}{L} \sum_{i=-\infty}^{\infty} \sum_{j=-\infty}^{\infty}
I_{j-i}(2 \tilde D_M t) e^{-2\tilde D_M t} \mathcal{I}_i(\alpha) \mathcal{I}_j(\alpha) 
\label{Cunif}
\end{eqnarray}
with
\begin{equation}
\mathcal{I}_k(\alpha) = \int_{ka}^{(k+1)a} \mathbf{1}_{[-L/2,L/2]}(x-\alpha) \; {\rm d}x.
\end{equation}
This quantity is in general equal to $a$ or 0, except when $k$ is the index of a box only partially covered by the observation profile (see Figure~\ref{defs}). The prefactor $1/L$ in Eq.~(\ref{Cunif}) is the product of $\sqrt{S_{\mathrm{eff}}}$ by the square of the normalization of the profile, namely $\left( 1/L \right)^2$. 

To go further, we need introducing some notations (Figure~\ref{defs}). From now, all lengths are expressed in units of $a$ (thus $a=1$ and $\tilde D_M = D_M$). We set $L=\ell+\varepsilon$ with $\ell$ an integer and $\varepsilon \in [0,1[$. The shift $\alpha \in [0,1[$ is the distance of the left-hand-side extremity of the segment to the closest obstacle on its left. We also introduce the integer $l_{max}$, indicating the number of boxes fully covered by the observation segment: $l_{max} = \ell$ if $\alpha+\epsilon<1$ and $l_{max} = \ell-1$ if $\alpha+\epsilon\geq 1$. In addition, two end-boxes are generally only partially covered by this observation segment. Their lengths are $\beta=1-\alpha$ and $\gamma=L-l_{max}-\beta$. 

Setting $z=2 D_M t$, and collecting the Bessel functions of same index, the correlation function is re-written as 
\begin{eqnarray}
C(t,L,\alpha) & = & \frac{e^{-z}}{L} \left[ (l_{max} + \beta^2  + \gamma^2) I_0(z) + 2 \sum_{j=1}^{l_{max}-1}
(l_{max} -j) I_j(z) \right. \label{long:first}\\ \nonumber
 & + & \left. 2 (\beta + \gamma) \sum_{j=1}^{l_{max}} I_j(z) + 2 (1-\delta_{l_{max},-1})\beta \gamma
 I_{l_{max} + 1} (z) \right].
 \label{Cdef}
\end{eqnarray}
Now we can define the average of $C(t,L,\alpha)$ over all possible
shifts $\alpha$:
\begin{eqnarray}
\bar C(t,L) & = & \int_0^1 C(t,L,\alpha) \; {\rm d} \alpha \\ \nonumber
 & = &  \int_0^{1-\varepsilon} C(t,L,\alpha) \; {\rm d} \alpha + 
       \int_{1-\varepsilon}^1 C(t,L,\alpha) \; {\rm d} \alpha .
\end{eqnarray}
In the first (resp. second) integral above, $l_{max} = \ell$ (resp. $l_{max} = \ell-1$). After 
integration, one gets:
\begin{equation}
\bar C(t,L)  = e^{-z} \left[(1-\frac{1}{3L})I_0(z) + 2 \sum_{j=1}^{\ell -1} (1-\frac{j}{L}) I_j(z)
+(\frac{1}{3} + \varepsilon + \varepsilon^2 - \frac{\varepsilon^3}{3}) \frac{I_{\ell}(z)}{L}  
+ \frac{\varepsilon^3}{3} \frac{I_{\ell+1}(z)}{L}   
\right].
\label{Cint}
\end{equation}
Figure~\ref{ex:fcs} shows an example of this long-term approximation (lower curve): it perfectly follows the exact $\bar C$ (upper curve) in the region of interest $t > \tau_m$. Now we use the following large $z$ expansion~\cite{Abra} where $\mu=4j^2$:
\begin{eqnarray}
\label{approxBessel}
e^{-z} I_{j} (z) & =  & \frac{1}{\sqrt{2 \pi z}}  \left[  1 - \frac{\mu -1}{8z} + 
\frac{(\mu -1)(\mu-9)}{2! (8z)^2} - 
\frac{(\mu -1)(\mu-9)(\mu-25)}{3! (8z)^3}  
+ \ldots \right] \nonumber \\
 & = & \frac{1}{\sqrt{2 \pi z}} e^{-\mu/8z} \left[ 1 + 
 \frac{1}{2z}\left( \frac{1}{4} -  \frac{\mu}{8z} +  \frac{1}{3} \left( \frac{\mu}{8z} \right)^2    \right)    + O\left( \left( \frac{1}{2z} \right)^2\right)  \right] \nonumber \\
 & = & \frac{1}{\sqrt{2 \pi z}} e^{-\mu/8z} \left[ 1 + 
 \frac{1}{2z} \mathcal{P}\left( \frac{\mu}{8z} \right) + O\left( \left( \frac{1}{2z} \right)^2\right)  \right],
\end{eqnarray}
after re-summing the terms of order $O(1)$ and $O(1/z)$. We defined the polynomial
$\mathcal{P}(X)=\frac{1}{4} - X + \frac{1}{3} X^2$. We recall that $x$ is the adimensional  variable of order 1:
\begin{equation}
x = \frac{2\sqrt{D_M t}}{L} = \frac{\sqrt{2z}}{L}.
\end{equation}
Thus $2z= x^2L^2$ and $\mu/(8z) = j^2/(L^2x^2)$. Now
at the leading order in $1/z$, only the first two terms contribute in
Eq.~(\ref{Cint}) and
\begin{eqnarray}
\bar C(x) & \simeq &  \frac{\sqrt{2}}{\sqrt{\pi z}L} \sum_{j=0}^\ell (1-\frac{j}{L}) \exp(-\frac{j^2}{2z}) \nonumber \\ 
 & \simeq & \frac{\sqrt{2}}{\sqrt{\pi z}} \int_0^1 (1-u) \exp(-\frac{u^2}{x^2}) {\rm d} u,
\end{eqnarray}
setting $u=j/L$. Thus $\bar C (x) = \Gamma(x) + O(1/L)$ with
\begin{equation}
\Gamma(x) = {\rm erf}(1/x)-\frac{x}{\sqrt{\pi}} (1 -
\exp(-1/x^2)),
\label{C:x:def}
\end{equation}
as expected for FCS with an uniform observation profile in the case of free diffusion (see Section~\ref{free}), since at large time-scales, in other words at large observation length-scale $L$, the diffusion recovers a free character. The graph
of $\Gamma$ was displayed in Fig.~\ref{C:x:courbe}. The calculation of the
next term in the expansion of $\bar C$ in powers of $1/L$ is derived
in Appendix~\ref{app1}. It requires in particular to take into account the corrections due to the discrete summation over $j$. It turns out that the terms of order $1/L$ cancel and
that the first non-vanishing term is of order $1/L^2$:
\begin{equation}
\bar C(x) = \Gamma(x) + \frac{g(x)}{L^2} + O(1/L^3),
\end{equation}
with
\begin{equation}
g(x) = \frac{1}{\sqrt{\pi} x } \left[ - \frac{1}{12}
  \left(1-\exp(-1/x^2) \right) + \frac{1}{6x^2}
  \exp(-1/x^2)\right].
\end{equation}
Now the decorrelation time $t_{1/2}$ is defined by $\bar C(2 \frac{\sqrt{D_M t_{1/2}}}{L}) = 1/\sqrt{2}$ and $x_0$ by $\Gamma(x_0) = 1/\sqrt{2}$ for $x_0 \simeq 0.520$. We recall that it means that, at the leading order in $L$,
$t_{1/2} = \frac{x_0^2}{4 D_M} L^2$ (see Section~\ref{free}).  Owing to the work of
Wawrezinieck {\em et al.}~\cite{Wawre05}, we anticipate that
\begin{equation}
t_{1/2} = \frac{x_0^2}{4 D_M} (L^2 - \sigma) + O(1/L)
\end{equation}
where $\sigma$ is the shift in $L^2$ observed on $t_{1/2}$ vs $L^2$
plots in numerical simulations ($\sigma$ is also in units of $a^2$). In
other words,
\begin{equation}
x_{1/2} \equiv 2 \frac{\sqrt{D_M t_{1/2}}}{L} = 
x_0 \left( 1 - \frac{\sigma}{2L^2}  \right) + O(1/L^3).
\end{equation}
Now 
\begin{eqnarray}
\bar C(x_{1/2}) & = & \Gamma(x_{1/2}) + \frac{g(x_{1/2})}{L^2} + O(1/L^3) 
  \nonumber \\
 & = & \Gamma(x_0) + \Gamma'(x_0) (x_{1/2}-x_0) + \frac{g(x_0)}{L^2} + 
O(1/L^3).
\end{eqnarray}
The second order term vanishes because $\bar C(x_{1/2}) = \Gamma(x_0)$, thus
\begin{eqnarray}
\label{sigma} \sigma & = & \frac{2 g(x_0)}{x_0 \Gamma'(x_0)} \\
\label{sigmabis}& = &  \frac{1}{2x_0^2}\left[ \frac{1}{3} - \frac{2}{3x_0^2} \;
\frac{1}{\exp(1/x_0^2)-1} \right] \\
 & \simeq & 0.49918\ldots
\end{eqnarray}
This quantity having been calculated with a high-precision software,
it is close to, but definitely different from 1/2. However, the value $\sigma = 1/2$ 
is an excellent approximation for a square meshgrid and a square 
observation area.

\begin{figure}[ht]
\begin{center}
\ \epsfig{figure=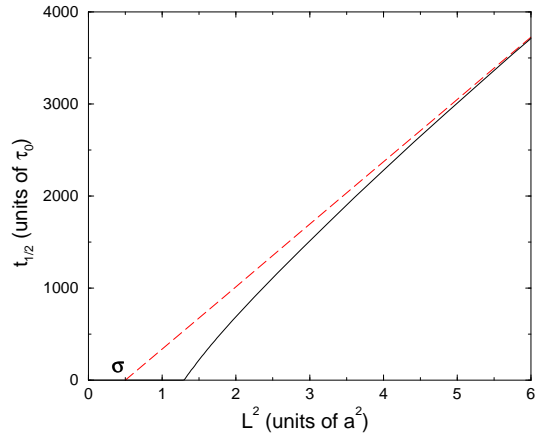,width=7cm} \
\end{center}
\caption{Decorrelation time, $t_{1/2}$, in arbitrary unit $\tau_0$, vs the observation domain area, $L^2$, for a uniform observation profile and $D_M = 10^{-4}a^2 \tau_0^{-1}$. The continuous line shows the numerical solution of $\bar C(t_{1/2},L)=1/\sqrt{2}$ with $\bar C$ given in Eq.~(\ref{Cint}). The dotted line is the result of the asymptotic expansion of $t_{1/2}$, valid at large $L$. It crosses the $L^2$ axis at $\sigma\simeq 0.5$. The numerical solution meets the $L^2$ axis at $L^2 \simeq  1.30$.}
\label{t1:2}
\end{figure}

We have shown in this section that, for a square uniform observation profile, up to corrections of order $1/L$ or higher,
\begin{equation}
\label{approx:uni}
t_{1/2} \simeq \frac{0.0676}{D_M} (L^2 - 0.499 \; a^2),
\end{equation}
where the length units are again expressed in real units (we recall that $a$ is the mesh parameter). This law is illustrated in Figure~\ref{t1:2}. The abscissa-shift $\sigma \simeq 0.499 \; a^2$ is the consequence of the difference between: (i) the correlation function $\bar C(t)$, when one takes carefully into account the effects of semi-permeable barriers on diffusion; (ii) and the same function $\Gamma(t)$ when one only takes into account the long-term effect of these barriers, which results in free diffusion with an effective coefficient $D_M$. This area-shift $\sigma$ amounts to a \emph{decrease} of the diffusion time $t_{1/2}$ with respect to the zeroth-order approximation $t_{1/2} \simeq 0.0676 \; L^2/D_M$. The main contribution to $\sigma$ comes from the term $1/3$ in the brackets of Eq.~(\ref{sigmabis}), the second term being approximately equal to 0.06. By inspection of Appendix~\ref{app1}, this term $1/3$ has 3 different contributions: $1/3 = 4(-1/12-1/6+1/3)$. The first two \emph{negative} contributions come respectively from the short-time corrections to the approximation of the Bessel functions (materialized by the polynomial $\mathcal{P}$, Eq.~(\ref{A5})) and the approximation of the sum on $j$ by an integral (Eq.~(\ref{A7})) .  The only \emph{positive} contribution to $\sigma$, that gives it its final \emph{positive} sign of interest here, come from the two extremities of the observation segment of length $L$ (segments f length $\beta$ and $\gamma$ in Fig.~\ref{defs}): there, the local in-box correlations fall off much more rapidly than the global correlations between boxes, which reduces the overall decorrelation time $t_{1/2}$. Beyond $\sigma$, the next-order corrections to $t_{1/2}$ can in principle be calculated by expanding the above calculations at order $1/L^3$. It would allow to estimate for which values of $L^2$ the approximation (\ref{approx:uni}) is valid. However, a rapid inspection of Fig.~\ref{t1:2} shows that the approximation is excellent whenever $L^2>6 \; a^2$ or $L>2.5 \; a$. 
At shorter $L^2$, the next-order corrections cannot be neglected and must be taken into account when exploiting experimental data. This issue will be addressed in the Discussion (Section~\ref{cl}), together with the additional effect of the short-term decorrelation neglected so far.

\subsection{Calculation of the short-term correlation function}
\label{short:uniform}

Our purpose here is to understand the short-time ($t < \tau_m$) behavior of the correlation functions, in particular the first fall-off of $C(t)$ observed in Figure~\ref{ex:fcs}. In contrast to the previous calculations, the short-term averages $\langle \cdot \rangle_{st}$ must be taken into account here. The interest of working with a uniform observation profile is that only end-boxes, of lengths $\beta$ and $\gamma$, will contribute at short times, which renders easier the understanding of underlying physics.

First of all, we consider the box of index $i$, the boundaries of which are approximated by perfectly reflecting ones. This approximation is correct at times $t < \tau_m \ll Ð\tau_e$ because exchange of matter between adjacent boxes is negligible. This box is observed with a uniform observation profile of intensity $1/L$ and length $\eta$, one extremity of which coincides with a box boundary, situated without loss of generality at $x=0$. This case includes both the two end-boxes (with $\eta = \beta$ or $\eta=\gamma$) and internal boxes that are entirely covered by the observation profile (with $\eta = a$). This observation profile is written $\iota(x)=\mathbf{1}_{[0,\eta]}(x)/L$. The total box filling, $W_i$, is constant, but can be different from 1 here since it fluctuates slowly at large times. The particle density at the position $x$ in the box, $\rho(x,t)$, is written $\frac{W_i}{a} \psi_i(x,t)$, this last function $\psi_i$ taking into account short-time local fluctuations in the box. It is a Poisson random variable of mean $\langle  \psi_i \rangle_{st}=1$ and mean square $\langle  \psi_i(x,t)\psi_i(y,t) \rangle_{st}=1+\frac{a}{W_i} \delta(x-y)$.

Like free diffusion (Section~\ref{free}), this case is tackled by Fourier analysis, which requires the calculation of the Fourier modes of the diffusion equation in this special geometry. We shall not detail these calculations here, which can be found in  text-books~\cite{Logan}, but simply give the final result. The short-term contribution of this box $i$ to $C(t,L,\alpha)$ is

\begin{equation}  
\label{box:correl}
\frac{L W_i^2}{a^2}  \int_0^a \mathrm{d}x \int_0^a \mathrm{d}y\;  \iota(x) \iota(y) \left[ \langle  \psi_i(x,s)\psi_i(y,s+t) \rangle_{st}  - \langle  \psi_i \rangle_{st} \langle \psi_i \rangle_{st}\right]= \frac{W_i}{L} \frac{\eta}{a} G_\eta(t),
\end{equation}
where
\begin{equation}
\label{G:eta}
G_\eta(t) = \frac{a}{\eta} \left[ 
\frac{\eta^2}{a^2} + \frac{1}{\pi^2} \sum_{n = -\infty \atop n \neq 0}^\infty \frac{1}{n^2}  \sin^2\left( n \pi \frac{\eta}{a} \right)
\exp \left( - \frac{D_\mu}{a^2} \pi^2 n^2 t \right) \right].
\end{equation}
The prefactor $L=\sqrt{S_{\mathrm{eff}}}$ in Eq.~(\ref{box:correl}) comes again from the overall normalization of $C(t,L,\alpha)$.  First of all, as anticipated in Eq.~(\ref{tau:mu}), these correlations fall off exponentially for times larger than $\tau_\mu = a^2/(\pi^2 D_\mu)$; $G_\eta(t)$ decreases from $1$ at $t=0$ to $\eta/a$ when $t \gg \tau_\mu$. If $\eta =a$, then $G_\eta(t)=1$ is constant. Internal boxes do not participate to the short-term decorrelation. In addition, consider two different boxes, with indices $i$ and $j$, $i \neq j$. Since inter-box currents vanish at the time-scale under interest, the short-term correlations, $\langle  \psi_i(x,s)\psi_j(y,s+t) \rangle_{st} - 1$, are also negligible. Therefore the only contribution to short-term decorrelation comes from the auto-correlation of both end-boxes. 

Now we must take the long-term average $\langle  \cdot \rangle_{lt}$, which amounts to replacing $W_i$ by $\langle  W_i \rangle_{lt}=1$. We set again $a=1$. The short-term contribution of both end-boxes to $C(t)$ is then
\begin{equation}
\label{final:short}
\frac{\beta G_\beta(t) + \gamma G_\gamma(t)}{L} = 
\left\{\begin{array}{lll}
  \displaystyle{\frac{\beta+\gamma}{L}}& \mathrm{if}&  t=0\\ 
  \phantom{a}\\
 \displaystyle{\frac{\beta^2+\gamma^2}{L}} & \mathrm{if}& t \gg \tau_\mu\\
 \end{array}\right. .
\end{equation}
The exact shape of this contribution appears in Figure~\ref{ex:fcs}, at times $t<\tau_m$(after integration over $\alpha$).
One can remark, from comparison of the algebraic expressions (\ref{long:first})~-- where $e^{-z} I_0(z) \simeq 1$ since $z = 2 \tilde D_M t = 2 t / \tau_e \ll 1$~-- and (\ref{final:short}), or from direct inspection of the figure, that our short- and long-term correlations functions coincide at intermediate times $t \sim \tau_m$. This is an additional validation of our approach.

\subsection{Calculation of the long-term correlation function for a Gaussian observation profile}

Here we focus again on the long-term correlation function, but in the Gaussian case. The short-term one can be treated in the same way as in the square case, with the additional complication that all boxes will contribute to it. The final result will be that the short-term contribution will also vanish exponentially for times larger than $\tau_\mu$, since this is a property associated with the boxes of size $a$ and not with the precise shape of the observation profile. By contrast, the numerical pre-factors of the decaying exponentials will be different from those of Eq.~(\ref{G:eta}) (the squared sines). However our main goal in the present paper is to tackle the calculation of the shift $\sigma$, which only is of practical interest~\cite{Wawre05}. This requires the sole long-term correlation function $\bar C$. Its calculation follows the same route as in the previous square case, in Section \ref{square:long}. We shall see that the dominant contribution comes now from the approximation of the discrete sum over boxes by an integral. The long-term diffusion is enhanced by the rapid diffusion in boxes that rapidly equilibrates the local particle densities.

The normalized one-dimensional Gaussian illumination intensity profile is now (see Section~\ref{profiles}):
\begin{equation}
I(x) = \frac{\sqrt{2}}{\sqrt{\pi} w}  \exp\left(  -\frac{2x^2}{w^2} \right),
\end{equation}
where the waist, $w$, plays the same role as the length $L$ in the previous uniform case. Again, this profile can be centered anywhere relatively to the meshgrid.
The function $C(t,w,\alpha)$ is defined as in the square case (Eq.~(\ref{Cunif})):
\begin{equation}
\label{Cgauss}
C(t,w,\alpha) = \frac{2}{\sqrt{\pi}\omega} \sum_{i=-\infty}^{\infty} \sum_{j=-\infty}^{\infty}
I_{j-i}(2 D_M t) e^{-2D_M t} \mathcal{I}_i(\alpha) \mathcal{I}_j(\alpha) 
\end{equation}
with
\begin{equation}
\mathcal{I}_k(\alpha) = \int_k^{k+1} \exp \left(  - \frac{2(x-\alpha)^2}{w^2}  \right) {\rm d}x.
\end{equation}
The prefactor $\frac{2}{\sqrt{\pi}\omega}$ is the product of $\sqrt{S_{\mathrm{eff}}}$ by the square of the normalization of the Gaussian, $\left( \frac{\sqrt{2}}{\sqrt{\pi} w} \right)^2$. We first integrate on the position of the center $\alpha$ of the intensity profile: 
\begin{equation}
\label{barCGauss}
\bar C(t,w) = \int_0^1 C(t,w,\alpha) \; {\rm d}\alpha = \frac{2}{\sqrt{\pi}\omega} \sum_{i=-\infty}^{\infty} \sum_{j=-\infty}^{\infty}
I_{j-i}(2 D_M t) e^{-2D_M t}  \mathcal{K}_{ij},
\end{equation}
where 
\begin{equation}
\label{Kij}
\mathcal{K}_{ij} =\int_0^1 \mathcal{I}_i(\alpha) \mathcal{I}_j(\alpha)\;  {\rm d}\alpha  
= \int_i^{i+1}  {\rm d}x \int_j^{j+1} {\rm d}y \int_0^1  {\rm d}\alpha \; \exp \left(
- \frac{2}{w^2}  [(x-\alpha)^2 + (y-\alpha)^2]  \right) .
\end{equation} 
Now we develop the exponential in powers of $1/w$ and $\alpha$: $(x-\alpha)^2 + (y-\alpha)^2= \frac{1}{2}(x-y)^2 + 2 [\alpha-\frac{1}{2}(x-y)]^2$, thus
\begin{eqnarray}
\label{sum:alpha}
\exp \left(
- \frac{2}{w^2}  [(x-\alpha)^2 + (y-\alpha)^2]  \right)  & = &
\exp \left(
- \frac{1}{w^2}  [(x-y)^2 + (x+y)^2]  \right)  \nonumber \\ 
 & & \times \left[
1 + 4 \frac{x+y}{w^2} \alpha + 4 \frac{1}{w^2} \left( 2 \frac{(x+y)^2}{w^2}-1 \right) \alpha^2
\right] + O(1/w^3).
\end{eqnarray} 
Here and in the sequel of this section, one must keep in mind that the calculation will end with a summation over $i$ and $j$ (Eq.~(\ref{Cgauss})). Therefore, the term in the square bracket proportional to $(x+y)$ vanishes when summing over $i$ and $j$, by parity. In addition, the exponentials in Eq.~(\ref{sum:alpha}) are Gaussian weights of the order of $\exp(-(i+j)^2/w^2)$ or  $\exp(-(i-j)^2/w^2)$. Thus, when summing over $i$ and $j$, any factor of order $(i+j)^n$ or $(i-j)^n$ will have a final contribution of order $w^n$ that must be taken into account when developing any expression in powers of $1/w$. This is precisely the case for the factor $(x+y)^2$ in the previous expression, which will eventually give a contribution of order $w^2$ because $x \approx i$ and $y \approx j$ in Eq.~(\ref{Kij}). For this reason, the third term in the bracket, proportional to $\alpha^2$, also vanishes at order $1/w^2$ when summing over $i$ and $j$. Consequently, at order $1/w^2$ of interest here, there is no dependency of $C(t,w)$ on the observation profile position, $\alpha$. The integration over $\alpha$ to compute $\mathcal{K}_{ij} $ is trivial.
This remark, together with the forthcoming calculations, will confirm the numerical observation in Ref.~\cite{Wawre05} (Fig. 10B): at large $w$, decorrelation times $t_{1/2}$ become independent of $\alpha$.

The Bessel functions $I_{j-i}$ are treated again using Eq.~(\ref{approxBessel}) and
at order $1/w^2$, still using $z=2 D_M t$,
\begin{eqnarray}
\label{GaussianC}
\bar C(t,w) & = & \frac{\sqrt{2}}{\pi} \frac{1}{w\sqrt{z}} \sum_{i,j} \exp \left( -\frac{(j-i)^2}{2z} \right)
\left[ 1 + 
 \frac{1}{2z} \mathcal{P}\left( \frac{(j-i)^2}{2z} \right)   \right] \nonumber \\
  & & \times \int_i^{i+1} {\rm d}x \int_j^{j+1} {\rm d}y \; 
 \exp \left( - \frac{1}{w^2}  [(x-y)^2 + (x+y)^2]  \right)  .
\end{eqnarray}
At the leading order in powers of $1/w$, defining again $x=\sqrt{2z}/w$, we get $\bar C (x) = \Gamma(x) + O(1/w)$ with
\begin{eqnarray}
\label{C:x:gauss}
\Gamma(x) & =  & \frac{\sqrt{2}}{\pi} \frac{1}{w\sqrt{z}} 
\int_{-\infty}^{\infty} {\rm d}x \int_{-\infty}^{\infty} {\rm d}y \; 
\exp \left( -\frac{(y-x)^2}{2z}\right) \; \exp \left( - \frac{1}{w^2}  [(x-y)^2 + (x+y)^2]  \right) 
\nonumber \\
 & = & \frac{1}{\pi x w^2} \int_{-\infty}^{\infty} {\rm d}u \int_{-\infty}^{\infty} {\rm d}v \; 
 \exp \left( -\frac{u^2}{2z} - \frac{1}{w^2}  (u^2 + v^2)  \right) 
 \qquad \qquad \qquad [u=y-x; v=y+x]
  \nonumber \\
 & = & \left( \frac{1}{1+x^2} \right)^{1/2},
\end{eqnarray}
by Gaussian integration, as expected for FCS with a Gaussian observation profile (see Section~\ref{free}). In this case, $x_{0}=1$.

\begin{figure}[ht]
\begin{center}
\ \epsfig{figure=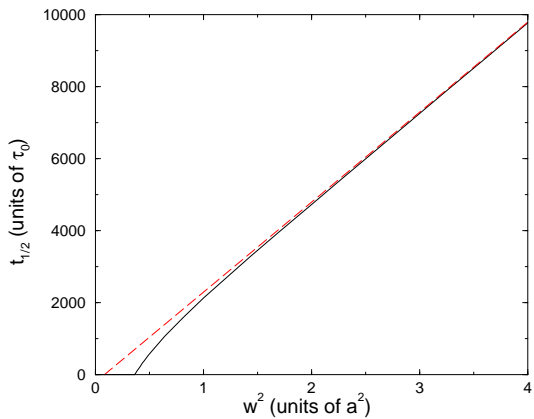,width=7cm} \
\end{center}
\caption{Decorrelation time, $t_{1/2}$, in arbitrary unit $\tau_0$, vs the observation domain area, $w^2$, for a Gaussian profile of waist $w$ and $D_M = 10^{-4}a^2 \tau_0^{-1}$. The continuous line shows the numerical solution of $\bar C(t_{1/2},w)=1/\sqrt{2}$. The dotted line is the result of the asymptotic expansion of $t_{1/2}$, valid at large $w$. It crosses the $w^2$ axis at $\sigma = 1/12 \simeq 0.0833$ (see text). The numerical solution meets the $w^2$ axis at $w^2 \simeq 0.362$.}
\label{t1:2Gauss}
\end{figure}

The corrections of order $1/w^2$ to $\Gamma(x)$ come from: (i) the term that is explicitly in  $1/(2z)=1/(x^2w^2)$; (ii) the discrete summation over $i$ and $j$, as in Eq.~(\ref{trapeze:error}). The case (i) is easily handled by Gaussian integrals. It gives a correction 
 $\frac{\sqrt{2}}{32} \frac{1}{w^2}$ at order $1/w^2$. 
The case (ii) requires a finer analytical investigation to be sure to track all the terms of the relevant orders. It is treated in Appendix~\ref{app2} and gives another correction 
 $-\frac{\sqrt{2}}{24} \frac{1}{w^2}$. One finally gets 
\begin{equation}
\bar C(x) = \Gamma(x) + \frac{g}{w^2} + O(1/w^3),
\end{equation}
with $g$ now independent of $x$:  
\begin{equation}
g = \frac{\sqrt{2}}{32}\ - \frac{\sqrt{2}}{24} = - \frac{\sqrt{2}}{96}.
\end{equation}
Finally, as in the previous uniform case,
\begin{equation}
\sigma = \frac{2 g}{x_{0} \Gamma'(x_{0})} = \frac{1}{12} \simeq 0.0833.
\end{equation}
This quantity can now be compared to the results obtained by numerical simulations by Wawrezinieck \emph{et al.}~\cite{Wawre05}, by relating it to the quantity $X_c^2$ computed there:
$X_c^2 \equiv \pi \sigma \simeq 0.262$ (in their case, the meshgrid parameter is $r=a/2$). Owing to the statistical errors inherent to any numerical method, this quantity had been over-estimated in previous publications and had been found to be larger than 1. However, the present analytical calculations confirm that the phenomenon pointed out by these numerical investigations enables the detection by FCS of underlying barrier meshgrids impeding long-term diffusion. We have also computed numerically the exact $t_{1/2}$ vs $L^2$ plots, as shown in  Fig.~\ref{t1:2Gauss}, by solving the equation $ 
\bar C(t_{1/2},L)=1/\sqrt{2}$, where $\bar C(t,L)$ is given in Eq.~(\ref{barCGauss}). They resemble those of Fig.~\ref{t1:2}.

One notices that in this case, the order-two approximation 
\begin{equation}
t_{1/2} \simeq \frac{1}{4D_M} (w^2 - a^2/12)
\end{equation} 
is excellent as soon as $w^2 > 2 \; a^2$. Below this limit, $t_{1/2}$ is sensibly smaller than its affine approximation and vanishes at $L^2 \simeq 0.362$. 

\section{Discussion}
\label{cl}

In this work, we have confirmed, by exact analytical arguments, previous numerical findings of P.-F. Lenne and his collaborators~\cite{Wawre05,Wenger07}: FCS at variable observation area gives a typical signature of two-dimensional diffusion hindered by a meshgrid of semi-permeable barriers. The decorrelation time $t_{1/2}$ is close to an affine function of the observation area $w^2$ and its experimental determination allows that of the sub-wavelength mesh parameter. This technique has indeed been shown in the same references to be applicable to the characterization of the diffusion properties of proteins embedded in the plasma membrane of live cells. In this case, the meshgrid can be constituted of cortical filaments of the cytoskeleton interacting with the cytoplasmic protuberance of proteins~\cite{Sheetz83,Kusumi93}.

For the sake of pedagogy, we have first presented our calculations in the case of  square, uniform, observation profiles before focussing on Gaussian ones. Conclusions are qualitatively similar in both cases. We recall that $w$ and $L/2$ (or $w^2$ and $L^2/4$) played essentially the same numerical role in the case of free diffusion. To what respect does this similarity survives on a meshgrid? First of all, by visual inspection, one can see that Figs.~\ref{t1:2} and~\ref{t1:2Gauss} are quite similar. Being equivalent to free diffusion, their long-time behavior is indeed identical, providing that one identifies $w^2$ and $L^2/4$. Besides, at shorter times, the ratio of $\sigma$'s is $0.499/0.083= 6 \approx 4$, and the ratio of points where the numerically calculated $t_{1/2}$ vanish is $1.30/0.362 = 3.59 \approx 4$. Therefore the correspondence $L/2 \leftrightarrow w$ is essentially preserved when going from the uniform profile to the Gaussian one. To this respect, there is another interest in having studied the square uniform case: circular, uniform profiles cannot been tackled within our formalism. But we expect them to be in-between the Gaussian ones (which have the circular symmetry), and the square ones (which possess the uniform character), and thus to resemble the two previous ones.

Now, we restrict the discussion to Gaussian profiles of experimental interest. We have proven that the decorrelation time $t_{1/2}$ (at which the two-dimensional correlation function gets reduced by a factor 2), displays an affine behavior,
\begin{equation}
t_{1/2} = \frac{1}{4D_M} (w^2 - a^2/12),
\end{equation} 
provided that the waist, $w$, and the mesh parameter, $a$, satisfy $w^2 > 2 \; a^2$. As compared to the numerical simulations previously mentioned, this work provides an accurate value of the parameter $\sigma$ equal to $1/12$ in this equation. The numerically calculated value was over-estimated by a factor larger than 4, thus \emph{a priori} leading to an under-estimate of the mesh parameter $a$ by a factor 2. 

The former over-estimate of this parameter $1/12$ does not go into the right direction for a good experimental extraction of the mesh size $a$. Indeed~\cite{Wawre05,Wenger07}, to have a good estimate of the shift $\sigma$, one has to go down to observation length-scales only slightly above $\sqrt{\sigma} \simeq a/3$. When working at laser wavelength $\lambda$, this implies $a/3 \gtrsim \lambda/2$, i.e. $a \gtrsim 600$~nm, while in general mesh sizes are supposedly not larger than 100~nm~\cite{Sheetz83,Saxton05,Kusumi05}.
This problem can be bypassed by employing optical techniques overcoming the Rayleigh limit, such as nanometric apertures~\cite{Wenger07} or Stimulated Emission Depletion (STED)~\cite{Hell94}, both of which \emph{a priori} enable one to go down to deci-nanometric length-scales. In the latter case, the profile is also close to Gaussian~\cite{Hell94}. But another complication arises when working at this scale: near $\sigma$, $t_{1/2}$ is not an affine function of $w^2$, as shown in Figure.~\ref{t1:2Gauss}. Nevertheless, it can reasonably be approximated by an affine function that now crosses the $w^2$ axis at $w^2 \simeq 0.362$ (corresponding to $X_c^2 \equiv \pi \; 0.362 \simeq 1.14$, with the notations of Refs.~\cite{Wenger07,Wawre05}). It is this value of $X_c^2$ that must be used for these small values of $L^2$. This last argument explains the value $X_c \approx 1$ found numerically in these references, where $t_{1/2}$ is fitted linearly without taking into account the curvature of the numerical curve observed in Figure.~\ref{t1:2Gauss}, finally re-conciliating numerical and analytical approaches.

We should also emphasize the following important points. Our calculations rely in great part on the effect of the first fall-off of $C(t)$ due to microscopic decorrelation, around $\tau_\mu$. The calculation of $t_{1/2}$ takes into account this fall-off because it reduces the value of $t_{1/2}$ with respect to free diffusion. But to see this effect on experimental data, one must be sure that the FCS curve $C(t)$ is acquired on a time range, $[t_{min},t_{max}]$, that does include this fall-off. In other words, to calculate when $C(t)$ is equal to 1/2, one must know the reference values $C(t)=0$ at large times and $C(t)=1$ at short times. Thus one must choose $t_{min}$ strictly lower than $t_\mu$. Diffusion coefficients $D_\mu$ are smaller than 0.1~$\mu$m$^2$s$^{-1}$ for membrane proteins in live cells, and $a$ is larger than $30$~nm, as discussed just above. Thus $\tau_\mu$  is much larger than 1~ms. A value of $t_{min}$ of 0.1~ms should be sufficient to ensure that short-term decorrelations are correctly taken into account in all live-cell plasma membranes.

Furthermore, fluorophore photo-physics are responsible for additional noise in FCS profiles at very short times. The most common cause for flickering in the fluorescence intensity is the transition of dyes to their first excited triplet state, where they stay a relatively long time before relaxing to their ground state~\cite{Schwille00}. If the contribution of this effect is superimposed to the short-time decorrelation addressed in the present study, the first fall-off will be over-estimated and the corrections to $t_{1/2}$ will contain a systematic error. However, this noise decorrelates after a few $\mu$s~\cite{Schwille00}, and it will not interfere with the mechanisms of interest here if one takes $t_{min} \approx 0.1$~ms. But this remark shows that one must be sure that all sources of noise specific to fluorophores are completely decorellated at $t_{min}$ in order to analyze correctly the diffusion on a meshgrid. This remark should help experimentalists to orient their fluorophore choice.

In the previous section, we have also noticed that as far as the Gaussian observation profile is concerned, the dependency of $t_{1/2}$ on the observation profile position $\alpha$ vanishes. At the order of $1/w$ considered here, which is adapted to study the large $w$ behavior of $t_{1/2}$, $\sigma$ does not depend on $\alpha$. This point had already been observed in numerical simulations~\cite{Wawre05,Wenger07}. However, if we had worked at a higher order in $1/w$, we would have found that $\sigma$ has corrections of order $1/w$ or higher that do depend on $\alpha$. Indeed, at small $w$, it is clear on the simulation results that the plots of $t_{1/2}$ vs $w^2$ depend on $\alpha$. By contrast, in the case of a uniform profile, if $\alpha$ is fixed, then $t_{1/2}$ depends on $\alpha$ (calculations not shown) because $g_{\alpha}(x)$ depends explicitly on $\alpha$. We have chosen in our calculation to average over $\alpha$ before calculating $\sigma$: $g(x) = \int_0^1 g_{\alpha}(x) \; \mathrm{d}\alpha$. At the experimental level, this corresponds to averaging FCS correlation profiles over realizations (at fixed $w$) before extracting $t_{1/2}$. However, we could have chosen instead to calculate $t_{1/2}(\alpha)$ (or equivalently $\sigma_{\alpha}$) for each value of $\alpha$ before averaging $t_{1/2}(\alpha)$ over the different values of $\alpha$. This would mean experimentally that one calculates $t_{1/2}$ for each realization before averaging over realizations. But $\sigma_{\alpha}$ is linear in $g_{\alpha}(x_0)$ (Eq.~(\ref{sigma})) and $\int_0^1 \sigma_{\alpha} \; \mathrm{d}\alpha =  2 \int_0^1 g_{\alpha}(x) \; \mathrm{d}\alpha / (x_0 \Gamma'(x_0))= \sigma$. Thus the overall result does not depend on the way of averaging over realizations.

Another issue that we have not debated so far is the contribution of the confined microscopic diffusion in boxes. If $D_M/D_\mu$ is very small, then $\tau_\mu \sim a^2/D_\mu$ is also very small and the first fall-off in Figure~\ref{ex:fcs} is pushed away to the left. It will have a significant influence on $t_{1/2}$  only when $L^2$ is very close to $\sigma$. One then expects its contribution to be of order $L^2/D_\mu$. Since $D_M/D_\mu \ll 1$,  this contribution is completely negligible as compared to the one previously discussed. In contrast, if $D_M < D_\mu$, but their ratio is not vanishingly small (typically $D_M/D_\mu \sim 1/10$), one can still speak of diffusion on a meshgrid. But the short-term contribution cannot be neglected anymore in the calculation of $t_{1/2}$, as it is visible in numerical simulations~\cite{Wenger07,Wawre05}. This makes the estimation of $a^2$ more difficult, because the point $\sigma$ where the long-term contribution of $t_{1/2}$ crosses the $L^2$ axis is now hidden in the short-term contribution (see Refs.~\cite{Wenger07,Wawre05}). Even though it is outside the scope of the present paper to tackle this issue, the calculation of this effect is feasible in principle, starting from the full correlation function. In addition, the microscopic diffusion constant, $D_\mu$, is also theoretically accessible from experiments, because $\tau_\mu$ is \emph{a priori} measurable at the first fall-off of the correlation function when $L$ is close to $a$ (even if the FCS signal appears to be quite noisy for these time values). Once $D_\mu$ is known, if it appears to be large as compared to the measured value of $D_M$, then one can forget this issue. In the converse case, the knowledge of both $D_\mu$ and its theoretical contribution to $t_{1/2}$ enables one to subtract this contribution from $t_{1/2}$ and thus to extract a good estimate of $a$. This work remains to be done.

We must also discuss the fact that at the experimental level, the underlying meshgrid has no reason to be a periodic, square lattice. It is more likely to look like a random meshgrid with cells of variable sizes and shapes. One cannot expect from an analytical approach to solve such a complex issue. Only numerical calculations can provide reliable results. However, it is reasonable to expect that the mean mesh diameter will be very close to the size $a$ measured \emph{via} the procedure discussed here. Furthermore, the present work can be generalized to three-dimensional diffusion in a similar meshgrid of obstacles defining three-dimensional, semi-permeable cages (e.g. diffusion in a gel-like cytoskeletal mesh). One has simply to replace $C_{I(\mathbf{r)}}(t)=C^2(t)$ by $C_{I(\mathbf{r)}}(t)=C^3(t)$ in our analysis. The diffusion time, $t_{1/2}$, is then defined by $C_{I(\mathbf{r)}}(t_{1/2})=1/2\sqrt{2}$. All the argumentation remains valid.

Finally, the present technique is not able to tackle isolated domains where membrane proteins are temporarily trapped and diffuse more slowly, as described in Refs.~\cite{Wawre05,Wenger07}. It has been shown numerically by P.F. Lenne and his collaborators that in this case, $t_{1/2}$ still displays an affine behavior, but that $\sigma$ becomes negative. The reason for the failure of our approach is that the potential $U(x,y)$ cannot be written as a sum of one-dimensional potentials in this case. What happens if we write $U(x,y) = U_1(x) + U_1(y)$, where the $U_1(x)$ are one-dimensional, periodic potentials, alternating zones of low and high potentials and of slow and rapid diffusion coefficients, $D_\mu$? Then diffusers encounter bands, both in the $x$ and $y$ directions, where the potential is higher (or lower) and the diffusion slower. At the intersections of these bands, there are square regions where $U$ is even higher (or lower) and where diffusion is even slower. These isolated, square regions could mimic those perviously mentioned and this problem could be tackled from our point of view. However, the quasi-static box approximation is not valid anymore in this case because we want $b$ to be of the same order of magnitude as $a$, thus making unrealistic the condition $b\ll a$. Here also, work remains to be done in the future.


\bigskip

\noindent \textbf{Acknowledgments:} I am indebted to Pierre-Fran\c{c}ois Lenne for signaling me the problem solved in this paper. I also thank him and Laurence Salom\' e for fruitful discussions during the redaction of the manuscript. Bastien Loubet also contributed to this work as part of his master's project. Finally, I am grateful to J\'erome Wenger for his kind reading of the manuscript.

\appendix

\section{Expansion of $\bar C$ in powers of $1/L$ (uniform case)}
\label{app1}
In this appendix, we identify and calculate the different corrections of order $1/L$ or $1/L^2$ to Eq.~(\ref{C:x:def}). First of all, this expression~(\ref{C:x:def}) comes from the approximation of a discrete sum,
the first two terms in Eq.~(\ref{Cint}), by an integral:
\begin{eqnarray}
A & = & e^{-z} \left[I_0(z) + 2 \sum_{j=1}^{\ell -1} (1-\frac{j}{L}) I_j(z) \right] \\
   & = & \frac{\sqrt{2}}{\sqrt{\pi z}} \left[ \frac{1}{2} f(0) + \sum_{j=1}^{\ell -1} f(\frac{j}{L}) 
  + \frac{1}{2} f(\frac{\ell}{L}) \right] - \frac{\varepsilon}{\sqrt{\pi}xL^2} \exp(-1/x^2) + O(\frac{1}{L^3}) \label{A2}
\end{eqnarray}
where $f(v)=(1-v)\exp(-v^2/x^2) \left(1 + \frac{1}{L^2x^2}\mathcal{P}
\left( \frac{v^2}{x^2}\right) \right)$. 
The first term $B$ in Eq.~(\ref{A2}) is the approximation by the trapeze integration method of the integral:
\begin{equation}
B=\frac{\sqrt{2} L}{\sqrt{\pi z}} \int_0^{\ell/L} f(v) {\rm d}v =
\frac{2}{x\sqrt{\pi}} \int_0^1 f(v) {\rm d}v -
\frac{1}{\sqrt{\pi}xL^2} \exp(-1/x^2) \varepsilon^2 + O(\frac{1}{L^3}).
\end{equation}
Thus $B = B_0 
- \frac{1}{\sqrt{\pi}xL^2} \exp(-1/x^2) \varepsilon^2 +  R$
with 
\begin{eqnarray}
B_0 & = & \frac{2}{x\sqrt{\pi}} \int_0^1 f(v) {\rm d}v \\
 & = & \Gamma(x) + \frac{1}{L^2} \frac{1}{x^2} \int_0^1 (1-v) \exp(-v^2/x^2)  
 \;  \mathcal{P}\left( \frac{v^2}{x^2}\right)  {\rm d}v \nonumber \\
 \label{A5} & = & \Gamma(x) + \frac{1}{\sqrt{\pi}xL^2} \left[ \frac{1}{12} 
 + \exp(-1/x^2) \left(- \frac{1}{12} + \frac{1}{6x^2} \right) \right]
\end{eqnarray}
and the residue, $R$, is given by the standard error of the trapeze method on each 
interval $[j/L,(j+1)/L]$:
\begin{eqnarray}
\label{trapeze:error}
R & = & \frac{2}{x \sqrt{\pi}} \frac{1}{12 L^2}\frac{1}{L} \sum_{j=0}^{\ell -1} f''(\frac{j}{L})  + O(\frac{1}{L^3})
\nonumber \\
 \label{A7}& = & \frac{1}{6 \sqrt{\pi} x L^2} \int_0^1 f''(v){\rm d}v + O(\frac{1}{L^3})\\
 & = & \frac{1}{6 \sqrt{\pi} x L^2} (1-\exp(-1/x^2)) + O(\frac{1}{L^3}) \nonumber. 
\end{eqnarray}
Finally, the three remaining terms in the brackets of Eq.~(\ref{Cint}) of
order $1/L$ contribute to $\bar C(x)$ at the order $1/L^2$:
\begin{eqnarray}
 & & e^{-z} \left[ - \frac{1}{3L} I_0(z) +\frac{1}{L} \left( \frac{1}{3}
+ \varepsilon + \varepsilon^2 - \frac{\varepsilon^3}{3} \right) I_{\ell}(z)
+ \frac{\varepsilon^3}{3} I_{\ell+1}(z) \right] \nonumber \\
 & = & \frac{1}{\sqrt{\pi} x L^2} \left[ - \frac{1}{3} (1-\exp(-1/x^2))
+ (\varepsilon + \varepsilon^2)\exp(-1/x^2)\right]+ O(\frac{1}{L^3}).
\end{eqnarray}
Thence
\begin{equation}
\bar C(x) = \Gamma(x) + \frac{1}{\sqrt{\pi} x L^2} \left[ - \frac{1}{12}
  \left(1-\exp(-1/x^2) \right) + \frac{1}{6x^2}
  \exp(-1/x^2)\right] + O(\frac{1}{L^3}).
\end{equation}

\section{Corrections of order $1/w^2$ due to the discrete summation in 
Eq.~(\ref{GaussianC}) (Gaussian case)}
\label{app2}

In this appendix, we calculate the corrections of order $1/w$ and $1/w^2$ to $\bar C(x)
= \Gamma(x)$, in the Gaussian case, due to the discrete sum in
\begin{equation}
D(t,w)  =  \frac{\sqrt{2}}{\pi} \frac{1}{w\sqrt{z}} \sum_{i=-\infty}^\infty\sum_{j=-\infty}^\infty  \exp \left( -\frac{(j-i)^2}{2z} \right)  \int_i^{i+1} {\rm d}x \int_j^{j+1} {\rm d}y \; 
 \exp \left( - \frac{1}{w^2}  [(x-y)^2 + (x+y)^2]  \right).
\end{equation}
This expression is that of $\bar C(t,w)$ (Eq.~(\ref{GaussianC})) where all terms being explicitly of order $1/w^2$ or $1/z$ have been omitted. However, the summation cannot be replaced by an integral without taking care of the ensuing errors because they will also appear to be of order $1/w^2$. We set again the new variables $u=y-x$ and $v=y+x$, and
\begin{equation}
D(t,w)  =  \frac{1}{2}\frac{\sqrt{2}}{\pi} \frac{1}{w\sqrt{z}} \sum_{i,j} \exp \left( -\frac{(j-i)^2}{2z} \right)  \int_{i+j}^{i+j+2} {\rm d}v 
\int_{j-i-1+|v-(i+j+1)|}^{j-i+1-|v-(i+j+1)|} {\rm d}u \; 
 \exp \left( - \frac{u^2}{w^2} \right)\exp \left( - \frac{v^2}{w^2} \right).
\end{equation}
The $1/2$ prefactor comes from the Jacobian of the substitution of variables. First we expand the integral over $u$ in powers of $1/w$:
\begin{eqnarray}
 & & \int_{j-i-1+|v-(i+j+1)|}^{j-i+1-|v-(i+j+1)|} {\rm d}u \; 
 \exp \left( - \frac{u^2}{w^2} \right) \nonumber \\
 & = & 2 \exp \left( - \frac{(j-i)^2}{w^2} \right)
 \left[  (1 - |v-(i+j+1)|) + \frac{1}{6w^2} (1 - |v-(i+j+1)|)^3 
 \right] + O(1/w^4).
\end{eqnarray}
The next step consists now of integrating over $v$ while keeping track of the relevant terms in the expansions in powers of $1/w$. Tedious but straightforward calculations lead to 
\begin{eqnarray}
\label{Ouf}
D(t,w)  & = & \frac{\sqrt{2}}{\pi} \frac{1}{w\sqrt{z}} \sum_{i,j} \exp \left( -\frac{(j-i)^2}{2z} \right)  
 \exp \left( - \frac{1}{w^2}  [(i-j)^2 + (i+j)^2]  \right) \nonumber \\
  & & \times \left[  1 + \frac{(i+j+1)^2}{3w^2} - \frac{1}{6w^2}
  + \frac{1}{6w^2} (-1 + \frac{(j-i)^2}{w^2})  \right] + O(1/w^3).
\end{eqnarray}
Following the same route, we expand
\begin{equation}
\label{ }
E(t,w) = \frac{\sqrt{2}}{\pi} \frac{1}{w\sqrt{z}} \sum_{i,j}   \int_i^{i+1} {\rm d}x \int_j^{j+1} {\rm d}y \; \exp \left( -\frac{(x-y)^2}{2z} \right)
 \exp \left( - \frac{1}{w^2}  [(x-y)^2 + (x+y)^2]  \right).
\end{equation}
We get an expression comparable to $D(t,w)$, except that in the last term of the square bracket of Eq.~(\ref{Ouf}), both $w^2$ must be replaced by $\sigma^2$, where $\sigma^2 \equiv \left( \frac{1}{2z} + \frac{1}{w^2} \right)^{-1}$ is of the order of $w^2$.
Now $E(t,w)$ is a mere Gaussian integral that is equal to $\Gamma(x)$. By subtracting
from Eq.~(\ref{Ouf}) its equivalent for $E(t,w)$, we finally get:
\begin{equation}
D(x,t) = \Gamma(x) - \frac{\sqrt{2}}{24} \frac{1}{w^2} + O(1/w^3).
\end{equation}
These tricky calculations, as well as those of Appendix~\ref{app1}, have been checked using a numerical integration software.


\end{document}